\documentclass[12pt,a4paper]{article}
\usepackage[dvips]{color}
\usepackage{epsfig,color}
\usepackage{pstricks,graphicx,epsfig,color,amssymb,amsmath,amscd}
\usepackage{bm}
\usepackage[normalem]{ulem}
\usepackage{soul} 

\newcommand{\be}{\begin{eqnarray}}
\newcommand{\ee}{\end{eqnarray}}
\newcommand{\bi}{\bibitem}

\newcommand{\rar}{\rightarrow}

\definecolor{gold}{rgb}{0.89,0.78,0}
\definecolor{grn05}{rgb}{0,0.5,0}

\begin{document}

{
\title{\Huge{
{
Early formed astrophysical objects and\\
cosmological antimatter
} } }
\author{{
\Huge{{{A.D. Dolgov}}}}
\\[2mm]
\large {
{ { University of Ferrara, Ferrara 40100, Italy}} }\\
{ { { NSU, Novosibirsk, 630090, Russia, 
}} 
}}
\date{\Large{{
{{
}
}}}}

}
\maketitle

\abstract{
Astronomical observations of recent years show that the universe at high redshifts about ten is 
densely populated by the early formed objects: bright galaxies, quasars, gamma-bursters, and 
contains a lot of metals and dust. Such rich early formed varieties  have not been expected in 
the standard model of formation of astrophysical objects. There is serious tension between the 
standard theory and observations.We describe the model which naturally relaxes this tension and 
nicely fits the data. The model naturally leads to creation of cosmologically significant antimatter 
which may be abundant even in the Galaxy. Phenomenological consequences of our scenario and 
possibility of distant registration of antimatter are discussed.
}

\newpage

\section{Introduction \label{s-intro}}

Probably the most impressive prediction of quantum field theory is the prediction of antimatter.
{Antiparticles are predicted and observed in experiment,}
{but it is unknown if antimatter, i.e. antistars, antiplanets, antigalaxies exist  anywhere in the universe.} 
{{Presently cosmological antimatter (the real one not just antiparticles)
is actively searched for  by several groups} and more  sensitive detectors are
proposed for future observations.

In his famous work~\cite{dirac}, 
{P.A.M. Dirac,}  {predicted 
``with the tip of his pen'' a whole world of antimatter}
(not just a small planet). { He assumed initially that positively charged "electron" was proton!?} 
{Critics by Oppenheimer, who noticed that hydrogen would be unstable
if proton was a hole in negative continuum,
forced Dirac to conclude that "anti-electron" is a {\it new} particle with the same mass as the electron (1931).
At that time a hypothesis about a new particle was not as blameless as today.

Very soon after that, in 1933, 
Carl Anderson announced the discovery of the positron, which was awarded by 
Nobel prize in 1936. {According to the Anderson's  words: it was not difficult, simply nobody looked for that.}

Paul M. Dirac justly  got his Nobel prize in 1933 immediately after the Anderson's discovery.
In his Nobel lecture ``Theory of electrons and positrons'' (December 12, 1933) Dirac, in particular, said 
"If we accept the view of complete
symmetry between positive and negative electric
charge so far as concerns the fundamental laws of
Nature, we must regard it rather as an accident that
the Earth (and presumably the whole solar system)
contains a preponderance of negative electrons and
positive protons. It is quite possible that for some of
the stars it is the other way about, these stars being
built up mainly of positrons and negative protons. In
fact, there may be half of stars of each kind. The
two kinds of stars would both show exactly the same
spectra, and there would be no way of distinguishing
them by present astronomical methods."
However, as we see in what follows, now we know ways to distinguish a stars from an antistars by observations 
from the Earth.
{The spectra are not exactly the same,} {even if CPT is unbroken} and polarization of the electromagnetic radiation
or the type of emitted neutrinos/antineutrinos from supernovae can be  good indicators to a star made of antimatter. 

It is striking that 
{Dirac was the second person to talk about antimatter.}
{ In 1898, 30 years before Dirac and one year after the discovery
of electron (J.J. Thomson, 1897)
{Arthur Schuster} (another British physicist) conjectured that 
there might be other sign electricity, {\it antimatter}, and 
supposed that} { there might be 
entire solar systems, made of antimatter,}  {\it indistinguishable}   from ours.}

 In his paper "{Potential Matter. Holiday Dream"}~\cite{schuster} 
A. Schuster wrote
{``When the year's work is over and all sense of responsibility has 
left us, who has not occasionally set his fancy free to dream about 
the unknown, perhaps the unknowable?''} and later
{"Astronomy, the oldest and yet most
juvenile of the sciences, may still have some surprises in
store. May antimatter be commended to its case".}

{Schuster made one more fantastic and correct guess. He envisaged that
matter and antimatter are capable to annihilate and to produce {\it vast} 
energy.} Remember that this was done before Einstein established his famous relation between mass 
and rest energy, $E= m c^2$.

To eliminate antimatter in close vicinity to the Earth Schuster conjectured that 
matter and antimatter are gravitationally repulsive because antimatter has negative mass.  
Two such objects on close contact should have vanishing mass. 
As we know now, all that happened to be incorrect: matter and antimatter are gravitationally attractive and 
and masses of particles and antiparticles has the same sign, and, due to the CPT-theorem, their masses are equal
by magnitude. Presently we have other explanations why antimatter is not seen in our neighborhood. 
Moreover, as we argue in what follows,
abundant stellar-like antimatter objects may be hidden even in the Galaxy not far from the Earth. Other galaxies
may be also abundant with antimatter.

{It would be great if these two great prophets were right and anti-worlds do exist.}
{However, the standard, simplest approach to baryogenesis (Sakharov) leaves no room for anti-worlds.}
Still simple generalization of baryogenesis scenarios could lead to abundant anti-worlds.
{Very early formed  and/or too old astronomical objects, observed presently, especially during last few years, 
may signify an existence of antiworlds.}

{An observation of cosmic antimatter will give a clue
to physics of baryogenesis, to the mechanism of cosmological
C and CP breaking, and present an extra argument in  
favor of inflation.} Since generalized scenarios predict not just a single number of baryon to photon ratio,
$\beta = (N_B- N_{\bar B} )/N_\gamma$, but a whole function of space point,
{${\beta (x)}$,} the models are falsifiable.

\section{Traditional ways of antimatter search and observational bounds \label{s-bounds}}

There are essentially two main methods for search of cosmic antimatter used up to now. The first, older one,
is a registration of the the products of $e^+e^-$ and  $\bar p p$  annihilations into electromagnetic radiation. 
In the first case especially spectacular is 0.511 MeV line from $e^+e^-$-annihilation at rest, which is in fact observed
in the Galactic center, but its origin is still unclear.  Proton-antiproton annihilation would create $\sim 100$ MeV photons 
with rather wide spectrum resulting from the decays of pions originally created by the annihilation. 

A registration of excessive cosmic antiprotons and positrons at low energies could also be an indication for antimatter
but at the present time the flux of low energy antiprotons and positrons can be explained by the conventional mechanism.

These are of course indirect methods and even if excessive antiprotons, psitrons, or 
gamma-rays are observed, a further analysis is necessary to exclude traditional
mechanisms of their production.

Even less direct methods include an analysis of spectrum distortion of the  CMB (both frequency and angular fluctuations),
of the impact of antimatter on big bang nucleosynthesis (BBN), and f the large scale structure formation.

During the last decade there is a burst of experimental activity for  
direct search of cosmic antimatter, namely for the search of antinuclej in cosmic rays, which have negligible probability to be
secondary-produced in the universe.
There are presently the following missions for the direct search of cosmic antinuclej: \\
{1. BESS: Japanese Balloon Borne Experiment with Superconducting
Solenoidal Spectrometer.} \\
{2. PAMELA (Italian-Russian space mission): Payload for Antimatter 
Matter Exploration and Light-nuclei Astrophysics.}\\
3. AMS: AntiMatter Spectrometer (Alpha Magnetic Spectrometer),
CERN-MIT-NASA.\\
The following two, more sensitive, detectors are under discussion now:\\
{PEBS (Positron Electron Balloon Spectrometer,) 
search for cosmic positrons and antiprotons.}\\
GAPS (Gaseous Antiparticle Spectrometer), search for
X-rays from de-excitation of exotic atoms, 
{it may reach 2 orders of magnitude better sensitivity 
than AMS for the ratio ${\bar He /He}$. }

At the present time the following bound on the flux of the antihelium-4 in cosmic rays is found :
{BESS:} {${\bar He / He < 3\times 10^{-7}}$.}
More restrictive limits are expected:
{PAMELA:} { ${{\bar He}/{ He} < 3\times 10^{-8}}$;}
AMS-2: { ${\bar He / He < 10^{-9}}$} but not yet reported.
The fluxes above are normalized to {the observed flux of the cosmic ray helium which at 
${ E< 10}$ GeV/nuclei is equal to:}
\be
 {{ dN_{He}/dE = 10^{2} /{\rm m^2/str/sec/GeV}.}}
 \label{He-flux}
 \ee

{According to the calculations of ref.~\cite{Duperray} 
the expected fluxes of the  secondary produced anti-nuclei are the following.}
Anti-deuterium is produced in ${\bar p\, p}$
or ${\bar p\, He}$ collisions 
The predicted flux of anti-deuterium produced by this mechanism is
${{\sim 10^{-7} /m^{2}/ s^{-1} /sr/(GeV/n)},}$
i.e. 5 orders of magnitude lower than the observed
flux of antiprotons.
{The expected fluxes of secondary produced 
${^3\bar He}$ and ${^4\bar He}$ 
are respectively 4 and
8 orders of magnitude smaller than the flux of anti-D.}

{Observations and bounds, summary:}\\
{${ \bar p / p \sim 10^{-5}-10^{-4}}$, is observed and can
be explained by secondary production;}\\
observed  $He/p \sim 0.1$;
{the upper limit: ${\bar He / He < 3\times 10^{-7}}$}.\\
Theoretical predictions: ${ \bar d \sim 10^{-5} \bar p}$, 
${ ^3\bar He  \sim 10^{-9} \bar p}$,
${ ^4\bar He \sim 10^{-13} \bar p}$.\\
{From the upper limit on ${\bar{He}}$: the nearest
single antigalaxy should be at the distance larger than } 
{10 Mpc (very crudely).}

Production of antinuclej was studied at LHC at ALICE detector~\cite{alice}. According to its data
the production of an antinucleous
{with an additional antinucleon is suppressed only by factor about 1/300} which is by far milder than
the suppression factors presented above. Probably the difference is related to much higher energies at
which data of ALICE are taken. The events with such energies are quite rare in cosmic rays and do not have
significant impact on the total rate of the antinuclej production.

Traditional bounds of the amount of the cosmic antimatter derived from the fluxes of the 
 cosmic gamma rays did not change during the last few decades because they essentially valid to our galactic supercluster.
According to ref.~\cite{steigman-76} the nearest {anti-galaxy} could not be closer 
than {at $\sim$10 Mpc}, as follows from the analysis of  ${ \bar p p}$-annihilation in common intergalactic cloud.
{The raction of antimatter in colliding galaxies in Bullet Cluster} is bounded from above by 
{${ < 3\times 10^{-6}}$}~\cite{steigman-2008}.

Note that 
{CMB excludes {\it large} isocurvature fluctuations at distances ${ d> 10}$ Mpc} and 
{BBN excludes large ``chemistry'' fluctuations at ${ d> 1}$ Mpc.} It means in particular that in a simple model the antimatter 
domains or objects should be sufficiently small.

There is relatively recent bound on the amount of antistars in the galaxy~\cite{ballmoos}.
The  Bondi accretion of the interstellar gas to the surface of an antistar leads to the annihilation luminosity:
\be 
 L_\gamma \sim 3\cdot 10^{35} (M/M_{\odot})^2 v_6^{-3}
\label{L-gamma-anti}
\ee
It allow to put the limit {${N_{\bar *} / N_{*} <  4\cdot 10^{-5} }$} inside 150 pc from the Sun, where $M$ is the mass of the
star, $M_\odot$ is the solar mass, and $v_6$ is the star velocity in $10^6$ cm/sec.  

Note in conclusion that all
{the presented bounds are true if antimatter makes the 
same type objects as the {\it observed} matter ones.}
For example, as we show in what follows, compact fast antistars may be abundant
in the Galaxy but still escape observations. A recent analysis of bounds on different type of antimatter 
objects was performed in ref.~\cite{bdp}.  Earlier works on
cosmology with domains of matter and antimatter was discussed in the lectures~\cite{steck} , where a list of relevant references can be found.

\section{ Observations of antistars from the Earth \label{s-antistats}}

According to Dirac and Schuster cited in the Introduction,  antistars are indistinguishable from stars by astronomical
observations. The basis for that is the CPT theorem, though it was unknown to the both quoted persons. This theorem 
demands equality of the level positions in atoms and antiatoms, so spectral measurements at first sight do nor allow to 
distinguish if the light  is emitted by an atom or an antiatom. However, we will see below that this is not exactly true.

Of course if  CPT is broken, the masses of particles and antiparticles may be different, though it is not necessarily so. 
If the particle-antiparticle mass difference is non-zero, ${\Delta m \neq 0}$, the 
spectra of anti-atoms might be different in an irregular way,} not described by the redshift. However, even if ${\Delta m \neq 0}$, 
it must be extremely tiny. As it is argued in ref.~\cite{ad-van}, non-zero mass difference between, say, electron and positron or 
between proton and antiproton breaks gauge invariance of electrodynamics, which in turn leads to non-zero photon mass.
The upper bounds on the latter are extremely restrictive and they demand  very strong upper bounds on the mass difference
$\Delta m_{ee} \equiv | m_e - m_{\bar e} | <  2\cdot 10^{-17}  $ eV or even $\Delta m_{ee} <  4\cdot 10^{-29}  $ eV~\cite{ad-van}. 
So even if   $\Delta m_{ee} \neq 0$, its effect on the spectrum should be tiny.

{Below we assume that CPT is respected and show that there are 
still chances for distant observation of antiworlds.}

Since C and CP are broken, the widths of spectral lines of atoms and anti-atoms must be different~\cite{ad-ik-ar}.
Unfortunately the magnitude of the effect for hydrogen is extremely small, maybe because of an accidental cancelation:
\be 
\Delta \Gamma /\Gamma \sim 10^{-28}.
\label{Delta-Gamma}
\ee
{An amplification of the effect, in particular, in heavier atoms and in external magnetic or electric fields might be possible.}

Possibilities discussed in ref.~\cite{ad-van-miv} look much more realistic, 
One evident way is to establish radio communication with inhabitants of the stellar system under scrutiny.. 
It may help to determine if they are or are not antimatter 
creatures. It is possible but looks more proper for science fiction.

A perspective way, maybe even in not too distant future, is a study
{of stellar neutrinos versus antineutrinos,} 
{especially from energetic SN explosions.}
At the first stage of SN explosion 
{neutrinos} from the neutronization 
 reaction ${pe^- \rar n\nu}$ are emitted, while  an anti-SN would emit
 {antineutrinos}.

An interesting possibility is a search for antistars through  
detection of  polarized photons produced in weak interaction processes.
{These photons are longitudinally polarized and their energy can be well defined if they are
created in two body decays.}  Mono-energetic photons produced e.g. in 
${B\rar K^\star\gamma}$ should be left-handed because of dominance of 
${b\rar s\gamma}$ penguin transition with left-handed ${s}$-quark.
{However, one can hardly expect noticeable abundance of B-mesons in stars,}
{and most probably there is an equal amount of  ${B}$ and ${\bar B}$
mesons.}

{Stars with abundant strange quarks look more promising.} 
{The outer shell of such stars is populated by ${\Sigma}$-hyperons and 
the polarization of photons emitted in ${\Sigma^+ \rar p\gamma}$ decay could indicate
if the photons are emitted by hyperon or anti-hyperon.}
The polarization is large, {${\alpha = -0.76\pm 0.08}$} and the branching ratio is non-negligible
{${(1.23 \pm 0.05)\times 10^{-3}}$.}

Circular polarization of photons in the ${\gamma}$-transitions of nuclei was observed
in the terrestrial experiments:
{${P_\gamma = (4\pm 1) \cdot 10^{-5}}$ in ${^{175} Lu}$ transition with the emission of 395 keV photon,} 
$P_\gamma = -(6 \pm 1) \cdot 10^{-6}$ for 482 keV photon emitted in ${^{181}Ta}$ transition, and
$P_\gamma = (1.9 \pm 0.3) \cdot 10^{-5}$
for 1290 keV photon emitted in transition of ${^{41}K}$.
{Measurement of circular polarization of such photon lines is ideally suited for search of antistars.}

\section{Baryogenesis \label{bargen} }

{Prevailing point of view at the present time is that }
{all the universe is made only of matter, while
primeval antimatter is absent.}
{Still the fact that antimatter exists 
creates fundamental cosmological  puzzle:} 
{why the observed universe is 100\% dominated by 
matter?}
Antimatter exists but not antiworlds, why?
{The problem deepened because of approximate 
symmetry between particles and antiparticles.}
In fact before 1956, the common faith in exact 
C, P, and T symmetries looked unbreakable, but with passing years more and more out of these discrete 
symmetries were discovered to be broken.

{The puzzle of the  baryon asymmetry, i.e. of the observed
predominance of matter over antimatter 
is resolved by three Sakharov's principles~\cite{sakharov}:}\\
{I. Nonconservation of baryons.} \\
{II. Violation of symmetry between particles and antiparticles, 
i.e. C and CP.} \\
III. Breaking of thermal equilibrium.

{There is a plethora of baryogenesis scenarios
which can explain the single number:}
\be 
\beta_{observed} = \frac{N_B-N_{\bar B}}{N_\gamma} 
\approx 6\times 10^{-10}\,.
\label{beta-obs}
\ee
{The usual outcome: ${ \beta = const}$, }
{which makes it impossible to distinguish between models and} 
{does not leave space for cosmological antimatter.} But the situation is not so pessimisitc, a simple
and natural generalizations of the simplest models of baryogenesis allow for a lot of antimatter almost at hand.
Of all the models the particularly interesting one is the Affleck-Dime (AD) or supersymmetric baryogenesis~\cite{AD-BG}.
It is the only known scenario of baryogenesis which may create large baryon asymmetry,  ${\beta \sim 1}$, while
all other models are able to create only a very small $\beta$, so theoretical efforts are usually done to enhance
the asymmetry as much as possible to make it compatible with the observed value.

Supersymmetry  (SUSY) predicts existence of scalar bosons with non-zero baryonic number,  {${ B\neq 0}$.}
Such bosons may condense along {flat} directions of the quartic potential, which is a generic property of the potential
in such models:
\be
U_\lambda(\chi) = \lambda |\chi|^4 \left( 1- \cos 4\theta \right),
\label{U-of-lambda}
\ee
There can be also the mass term, ${ m^2 \chi^2 + m^{*\,2}\chi^{*\,2}}$, which has another set of flat directions:
\be
U_m( \chi ) = m^2 |\chi|^2} {\left[{ 1-\cos (2\theta+2\alpha)}  \right] .
\label{U-mass}
\ee
where ${ \chi = |\chi| \exp (i\theta)}$ and $ m=|m| \exp (\alpha)$.
If ${\alpha \neq 0}$, C and CP are  broken.

In grand unified  (GUT) SUSY models the baryonic number of $\chi$
is naturally non-conserved. It is expressed by the non-invariance of 
$U(\chi)$ (\ref{U-of-lambda}, \ref{U-mass}) with respect to the phase rotation, $\chi \rar \exp(i\theta) \chi$.

Initially (at inflation) ${\chi}$ is naturally away from the origin and when 
inflation is over, it starts to evolve down to the equilibrium point, ${\chi =0}$,
according to the equation of Newtonian mechanics for point-like particle:
\be
\ddot \chi +3H\dot \chi +U' (\chi) = 0.
\label{ddot-chi}
\ee
Baryonic charge of $\chi$:
\be
B_\chi =\dot\theta |\chi|^2
\label{B-chi}
\ee
is analogous to the mechanical angular momentum in the two-dimensional complex plane $[{Re\,}\chi,\,\,{ Im}\,\chi]$ . 
The decays of ${{\chi}}$ transferred the baryonic charge accumulated in the rotational motion of $\chi$ into that of quarks 
in B-conserving processes.

If ${ m\neq 0}$, the angular momentum, or what is the same the baryonic number, B, is generated by possibly different 
directions of the  quartic and quadratic valleys at low ${\chi}$.
{If CP-odd phase ${\alpha}$ is small but non-vanishing, both baryonic and 
antibaryonic regions are possible }
{with dominance of one of them.}
{Matter and antimatter domain may exist but globally ${ B\neq 0}$.} This is briefly the canonical version of the
AD-scenario which results in normal value of the baryon asymmetry with possible domains of antimatter of 
arbitrary sizes and values of $\beta$. Such a model of antimatter creation is strongly restricted by the observational
data.

{A minor modification of AD-scenario can lead to very early formation of compact stellar-type objects and
naturally to a comparable amount of anti-objects,}
{such that the bulk of baryons and (equal) antibaryons
are in the form of compact stellar-like objects, e.g. QSO or primordial black holes (PBH), formed in the regions with a huge 
baryon asymmetry,} 
{plus sub-dominant observed homogeneous baryonic background.}
The amount of antimatter may be comparable or even larger 
than of {\it known} baryons, but such ``compact'' (anti)baryonic objects
would not contradict  any existing observations.

Note that the astronomical data accumulated during the last few years demonstrate rich population of superheavy black 
holes in the very early universe, which indeed could be created in the frameworks of the modified AD-scenario,
but remain mysterious otherwise,

The necessary generalization of the AD-baryogenesis was suggested in ref.~\cite{AD-JS} and further developed in
ref.~\cite{ad-mk-nk}. To this end a general renormalizable coupling of the
Affleck-Dine field $ \chi$  to inflaton, $ \Phi$, has been introduced (the first term in the equation below), 
so the total potential of $\chi$ included the interaction with the inflaton, $\Phi$, gets the form:
\be 
U =  g | \chi |^2 (\Phi -\Phi_1)^2  +
\lambda | \chi|^4 | \,\ln\left( \frac{| \chi |^2 }{\sigma^2 } \right)
+\lambda_1 \left(\chi^4 + h.c.\right) + (m^2 \chi^2 + h.c.).
\label{U-of-Phi-chi}
\ee
The second term in this expression is the so called Coleman-Weinberg~\cite{CW} 
potential obtained by summation of one-loop radiative corrections.

When $\Phi$ is far from $\Phi_1$, the effective mass squared of $\chi$ is positive and large, so the gate to the flat
directions is closed. It is open only during relatively
short period, when ${\Phi \approx \Phi_1}$,
but still at inflationary epoch. Because of that  cosmologically small but possibly astronomically large 
bubbles with very  high ${ \beta}$ could be
created, occupying {a small
fraction of the universe volume,} while the rest of the universe has the normal baryon asymmetry
{${{ \beta \approx 6\cdot 10^{-10}}}$, created in the regions with small initial
${\chi}$}, occupying the bulk of the universe volume. Such process looks as phase transition of 3/2 order.
In the simplest version of the model the high-$\beta$ objects would have positive and negative baryon asymmetry
with practically equal probability. Interesting  anti-objects should be astronomically large, so inflation is
necessary and not too large to avoid problems with the
existing observations. So some, rather mild fine-tuning of the model parameter values is necessary.

The distribution of high baryon density bubbles over  mass (and size) has very simple log-normal form:
\be
\frac{dN}{dM} = C_M \exp{[-\gamma \ln^2 (M/M_0)]}
\label{dN-dM}
\ee
where ${C_M}$, ${\gamma}$, and ${M_0}$ are constant model dependent parameters.
{The spectrum is practically model independent, because it is essentially determined by the exponential expansion (inflation).} 

There are several (essentially two) periods when the density contrast between the regions with high baryonic density and those
with the normal low one arose:\\
1. Firstly, after formation of the  domains (bubbles) with high value of ${\chi}$ and correspondingly with huge baryon 
(anti-baryon) asymmetry
the initial density contrast between the high-B bubbles and the rest of the universe would be zero or very small, because quarks,
which carry baryonic number at this stage were massless. These are the so called isocurvature perturbations.
Somewhat later due to different equations of state inside and outside of the
domains (or what is the same of high-B bubbles)  some density contrast would evolve because the matter inside the bubbles were 
more non-relativistic due to higher amplitude of non-relativistic field $\chi$ inside.\\
2. The second period of perturbation rise is much more efficient. It takes place  
after the QCD phase transition { at ${ T\sim 100}$ MeV when quarks made non-relativistic protons.}
During this stage compact bubbles with high baryonic mass density would be formed. With properly chosen parameters such
bubbles could be primordial black holes (PBHs), or dense stellar like objects.
The PBH masses could range from a few ${M_\odot}$ up to ${10^{6-7} M_\odot}$. 

The model successfully explains quite a few mysterious facts observed by astronomers, with the data especially accumulated
during the recent few years. It presents a mechanism of superheavy black hole formation, and what is especially important, at very early time. 
Correspondingly it explains existence of quasars observed at high redshifts, $z \sim 10$, which is mysterious  in the standard model.
Moreover, the near-quasar space is enriched with metals. An explanation of this fact demands an unusual BBN (see below) or
an early supernova formation.
 High redshift gamma bursters also demand early cleated supernovae - this can  be simply arranged in the considered model.
Early formed superheavy BHs can be seeds of galaxy formation. In the accepted nowadays picture the process is supposed to be 
the opposite: first galaxies were formed and superheavy BHs in the center of practically all observed galaxies were created later by the
matter accretion to the galactic center. However, in the standard approach the rate of superheavy BH formation is too low to explain
their formation during the cosmological time. Moreover, several small galaxies are known with the central superheavy BH 
having mass similar by magnitude to the total mass of the galaxy. This surely does not fit the standard accretion picture.

{\it BBN in high-B environment.} 
If ${\beta \equiv \eta \gg 10^{-9} }$, light (anti)element 
abundances would be
anomalous: the clouds with high $\beta$ would contain much less (anti)deuterium and more (anti)helium, than the normal one.
{Evolved chemistry in the so early formed QSOs can be explained, at least to some extend,}
{by more efficient production of metals during BBN due to much larger ratio 
${\beta =N_B/ N_\gamma}$. }
The standard BBN essentially stops at ${^4}$He because of the very
small $\beta$. However, in the model considered here $\beta$ is
much larger than the canonical value, even being close or exceeding unity. BBN with anomalously high 
$\beta \gg \beta_{obs} = 6\cdot 10^{-10}$ 
was considered in ref.~\cite{bbn-anom} but still only up to $\beta \ll 1$. An extension to $\beta \sim 1$ would be very interesting.  

To summarize the model explains in the unique way the
formation of high-redshift supernovae, gamma-bursters, stars older than 
the universe observed in the Milky Way, cut-off BH distribution at $M < 6M_\odot$
in galaxies, and more. It is discussed in detail in the next section.

According to our estimates the compact anti-objects mostly survived in the early universe,
(especially if they are PBHs).
The model predicts an existence of a kind of early dense stars formed when the  
initial pressure outside was larger than internal pressure. Astrophysics of such objects may be quite unusual.
Such ``stars'' may evolve quickly and, in
particular, make early SNs, enrich the universe with heavy
(anti)nuclei and reionize the universe.

Phenomenology and bounds on compact antimatter objects and disperse anti-clouds was studied in the
papers~\cite{cb-ad,bdp}. The early formed stars and black holes are discussed in ref.~\cite{ad-sb}

\section{Mysteries in the sky \label{mysteries}}

On the background of quite successful description of the gross features of the universe the standard $\Lambda$CDM cosmology
encounters plenty of disturbing and interesting problems. These troubling data became especially prominent during the last
several years. In short, there is an avalanche of  astronomical discoveries of different kind of astrophysics objects
which could not be created and evolved  in the available cosmological time.
They are observed in the very early universe at redshifts $z \sim 10$ and in
the contemporary universe, e.g. in our Galaxy there are stars, which look older than the universe.
Rephrasing the famous words by Marcellus from "Hamlet" we can say: 
{"Something is rotten in the state of \st{Denmark} the Universe"}.
It is tempting to explain all these data by our model described in the previous section.
It is especially interesting that the suggested scenario predicts, as a byproduct abundant 
cosmological antimatter in close vicinity to the Earth. 

To see how old are the objects observed at different redshifts let us present the expression of the universe age $t_U$,
as a function of the cosmological redshift $z$:
 \be
t(z) = \frac{1}{H}\,\int_0^{{1}/({z+1)}} \frac{dx}
{\sqrt{1-\Omega_{tot} +{\Omega_m}/{x} + x^2\,\Omega_v } },
\label{t-U}
\ee 
where $\Omega_m$ is the fractional (in terms of the critical energy density) 
energy density of matter (i.e baryonic plus dark matter), 
and $\Omega_{v}$ is the fraction of the density of dark energy. The total fraction of the
cosmological energy density, $\Omega_{tot}$, is supposed to be unity (spatially flat universe), 
since the energy density of the CMB
photons and neutrino can be neglected during essential time interval in the integral.

According to the Planck data, the present day values of theses parameters are:
${\Omega_{tot} = 1}$, ${\Omega_m  = 0.317}$, and
${\Omega_v = 0.683}$. There is some tension between the values of
the Hubble parameter measured by Planck, ${ H_{pl} = 67.3}$ km/sec/Mpc and 
that measured by the traditional astronomical methods,
which can lead to $H$ as large as ${ H_{astr} =~74}$~km/sec/Mpc, see ref.~\cite{planck-prmtr} for discussion. 
Possibly both results might be correct and the difference in the measured values of $H$ emerges because the
Planck and the astronomical measurements are sensitive to  $H$ at different moments  of the cosmological history.
This e.g. can occur in a model where a fraction of dark matter is unstable and decays at some moment between the 
recombination and the present day~\cite{zb-ad-it}.

We present a few examples of the universe age in giga-years for different $z$. The first number corresponds to the 
Planck measured value of $H$, and the other, shorter one, in brackets to the larger astronomical value:
$ t_U \equiv t(0) = 13.8 \,(12.5.)$; $ t(3) =  2.14\, (1.94)$; ${ t(6) = { 0.93;}\, {(0.82)}}$; ${ t(10) = 0.47\,{(0.43)}}$; and
${ t(12) = { 0.37;}\,\, {(0.33)}}$.

\subsection{Strange phenomena in the Milky Way. \label{ss-MW}}

{\it Old stars in the Milky Way.} With an increased accuracy of the nuclear chronology the ages of several stars have 
been recently determined to be much older than expected. Below we quote a few recent results.

Employing thorium and uranium  abundances
in comparison with each other and with several stable elements {the age of
metal-poor, halo star BD+17$^o$ 3248 was estimated as~\cite{cowan}  
${13.8\pm 4}$ Gyr.
{For comparison the age of inner halo of the Galaxy is } {${11.4\pm 0.7}$ Gyr.}

The age of a star in the galactic halo, HE 1523-0901, was estimated to be 
about 13.2 Gyr\cite{frebe}.
{First time many different chronometers, such as the U/Th, U/Ir, Th/Eu and Th/Os ratios to
measure the star age have been employed.}

The metal deficient {\it high velocity} subgiant in the solar neighborhood
HD 140283  has the age ${14.46 \pm 0.31 }$ Gyr~\cite{H-Bond}
The central value exceeds the universe age by two standard deviations,
if ${H= 67.3}$ and ${t_U =13.8}$, while for  ${H= 74}$, and thus ${ t_U = 12.5}$ the excess is by nine
standard deviations. So this could really looks as a star older than the universe.

A possible explanation of this discrepancy could be an unusual initial chemical content of such stars.
Normally a pre-stellar cloud consists of 25\% of $He^4$ and 75\% of hydrogen. However, in our case 
the primordial nucleosynthesis is able to create much heavier elements, as is mentioned in the previous sections
or/and the early supernovae could enrich the interstallar gas with heavy elements, so the initial chemical 
content of some stars would be much different from the traditional one. Such stars could evolve to their
present state considerably faster than the usual ones. Of course this conjecture should be verified by
the stellar nucleosynthesis calculations.

{\it The mass distribution of BHs} observed in the Milky way demonstrates some peculiar features not understood by
the conventional theory. It is found that their masses are concentrated in the narrow range
${ (7.8 \pm 1.2) M_\odot }$~\cite{M-BH-narrow}.
This result agrees with another paper where
a peak around ${8M_\odot}$, a paucity of sources with masses below
 ${5M_\odot}$, and a sharp drop-off above
${10M_\odot}$ are observed~\cite{kreidberg}. Such facts are indeed very strange, if these BHs were formed
by the stellar collapse, as it is usually assumed.
However the log-normal mass distribution (\ref{dN-dM}) of the primordial black holes
 naturally explains such a form of the mass spectrum  of the stellar mass black holes in the Galaxy.
 {Astronomical data also indicate  a two-peak mass distribution of the PBHs and compact stars,} which is 
probably observed, but not explained up to now~\cite{farr}. Quoting this work:
"sample of black hole masses provides strong evidence of a gap between the maximum neutron star mass and the lower bound on 
black hole masses". These results also fit the mass distribution of our model, assuming that the neutron stars were mostly created
by the conventional mechanism.

\subsection{ Supermassive BHs and quasars in (near) contemporary universe\label{present-time} }

{it seems that every large galaxy and some smaller 
ones contain a central supermassive BH} with masses 
larger than { ${ 10^{9}M_\odot}$} in giant elliptical
and compact lenticular galaxies
and {${\sim10^6 M_\odot}$} in spiral galaxies like Milky Way.

The mass of BH is typically 0.1\% of the mass of the stellar bulge of galaxy~\cite{BH-bulge}
but some galaxies may  have huge  black holes: e.g. NGC 1277  has
the central BH  of  ${1.7 \times 10^{10} M_\odot}$, or ${60}$\% of its bulge mass~\cite{NGC1277}.
{This fact creates serious problems for the
standard scenario of formation of central supermassive BHs by accretion of matter in the central 
part of a galaxy.}
{An inverted picture looks more plausible, when first a supermassive black hole was formed and 
attracted matter serving as seed for subsequent galaxy formation.}

More similar examples can be found in ref.~\cite{khan-2015}. As the authors say,
although supermassive black holes  correlate well with their host
galaxies, there is an emerging view that outliers exist.
 Henize 2-10, NGC 4889,
and NGC1277 are examples of super massive black holes  at least an order of magnitude more massive
than their host galaxy suggests. }
{The dynamical effects of such ultramassive central black holes is unclear. }

{A  discovery of an ultra-compact dwarf galaxy
older than 10 Gyr, enriched with metals, and probably with a massive black in its center} 
also seems to be at odds with the standard model~\cite{strader}.
The dynamical mass of the galaxy is ${2\times 10^8 M_\odot}$ and the radius is
${R \sim 24}$ pc, so its density is very  high.
The Chandra data reveal a variable central X-ray source with $L_X \sim 10^{38}$ erg/s that could be an active galactic nucleus 
associated with a massive black hole or a low-mass X-ray binary. Analysis of optical spectroscopy shows the object to be 
old, $ \gtrsim 10$ Gyr and of solar metallicity, with elevated [Mg/Fe] and strongly enhanced [N/Fe] that indicates light element 
self-enrichment; such self-enrichment may be generically present in dense stellar systems.

Recently an observation of a quasar quartet embedded in giant nebula was reported~\cite{quartet}
 in a survey for Lyman-α emission at redshift  ${z \approx 2}$. According to the authors,
it reveals rare massive structure in distant universe.
All galaxies once passed through a hyperluminous quasar phase powered by accretion onto a 
supermassive black hole. But because these episodes are brief, 
{quasars are rare objects separated by cosmological distances, so
the chance of finding a quadruple quasar is ${\sim 10^{-7}}$.} 
It implies that the most massive structures in the distant universe have a tremendous supply 
(${\sim 10^{11} M_\odot}$) of cool dense (${ n \approx 1/}$cm$\bm{^3}$) gas,
{in conflict with current cosmological simulations.}

\subsection{Young objects at  high redshifts  \label{ss-hi-z}}

There is a large "zoo" of evolved astronomical objects formed in the universe in surprisingly short times.
{Several galaxies have been observed at high redshifts,} with natural gravitational lens ``telescopes''.
There is a galaxy at {${z \approx 9.6}$} which was formed when the universe was approximately
0.5 Gyr old~\cite{zheng-2012}. 

Moreover a galaxy at {${z \approx 11}$} has been observed~\cite{coe} 
which was formed eariler 
than the universe age was { 0.41 Gyr} (or even shorter with larger H).

An observation of not so young but extremely luminous galaxy was recently reported by WISE~\cite{wise}:
{{${L= 3\cdot 10^{14} L_\odot }$; age: ${\sim 1.3 }$ Gyr.}
{{For creation of such galaxy the galactic seeds, or embryonic black holes, might be bigger than thought possible.}
As one of the authors of the paper, P. Eisenhardt, said : "How do you get an elephant?  One way is start with a baby elephant."
{The BH was already billions of ${M_\odot}$ , when our universe was only a 
tenth of its present age of 13.8 billion years.}


{Thus we smoothly come to another and even more striking example of early formed objects, i.e. to quasars observed at high z.}
A quasar with maximum {${ z = 7.085}$} has been observed~\cite{Mortlock}. 
It means that  the quasar was 
formed at {${t< 0.75}$ Gyr.} Its luminosity and mass are respectively {$L={6.3 \cdot 10^{13} L_\odot}$} 
and {$M={2 \cdot 10^9 M_\odot}$.}
{The quasars are supposed to be supermassive black holes (BH) }
{and their formation in such short time looks problematic by conventional mechanisms.}

As it is summarized in ref.~\cite{melia}
{"rapid emergence of high-z galaxies} 
so soon after the big bang 
may actually be in conflict with 
current understanding of how they came to be. This
problem is very reminiscent of the better known (and
probably related) {premature appearance of supermassive
black holes at ${z\sim 6}$.} It is difficult to understand how
{${10^9 M_\odot}$} black holes
appeared so quickly after the big
bang without invoking non-standard accretion physics
and the formation of massive seeds, 
{both of which are not seen in the local Universe."}

Very recently another monster was discovered~\cite{xue-2015}: 
"An ultraluminous quasar with a twelve billion solar mass black hole at redshift 6.30".
There is already a serious problem with formation mechanism of lighter and less luminous quasars
which is now multifold deepened with this new "creature".
As is stated in the paper,
about 40 quasars with ${z> 6}$ are known, each quasar containing BH with 
${M \sim 10^9 M_\odot}$.  {Such black holes,
when the Universe was less than one billion years old, 
present substantial challenges to theories of the formation and growth of
black holes and the coevolution of black holes and galaxies.} Now we have an order of magnitude
heavier one, $M \approx 10^{10} M_\odot $.
It has an optical and nearinfrared luminosity a few times greater than those of
previously known $z> 6$ quasars.

As we have already mentioned, the universe at $z \sim 10$ is quite dusty, for a recent analysis and references 
see the paper~\cite{dust}.
{The medium around the observed early quasars contains
considerable amount of ``metals''} (elements heavier than He). 
According to the standard picture, only elements up to ${^4}$He  { and traces of Li, Be, B}
were formed in the early universe by BBN, {while heavier elements were created
by stellar nucleosynthesis and} {dispersed in the interstellar space by supernova explosions.}
{If so, prior to or simultaneously with the QSO formation a rapid star formation should take place.}
{These stars could produce plenty of supernovae which enriched interstellar space   by metals.}

Moreover, observations of high redshift gamma ray bursters (GBR) also indicate 
{a high abundance of supernova at large redshifts.} 
{The highest redshift of the observed GBR is $z=9.4$} and there are a few more
GBRs with smaller but still high redshifts. 
{The necessary star formation rate for explanation of these early
GBRs } {is at odds with  the canonical star formation theory.}

On the other hand, our model may naturally lead to high density of compact stars, which in particular can be
early supernova created much earlier than $z=10$. 

\section{Conclusion \label{s-concl}}

The observational data at high redshifts, $z \sim 10$, which became available in the recent years show that the 
universe is populated by the early formed bright galaxies, quasars, gamma-bursters, and contains a lot of metals 
and dust. Such rich early formed varieties  have not been expected in the standard model of formation of 
astrophysical objects. There is serious tension between the standard theory and observations.

Probably most striking is existence of supermassive black holes both in the early and in the present day universe.
The standard scenario is that such black holes have been formed through accretion of matter to an initial solar mass black hole. 
However, it demands much more time than that available during all the cosmological history. This is especially troubling for the
supermassive black holes (quasars) created at the dawn of the universe at high $z$. 
The inverted picture when primordial BHs were first created and then served as seeds for the galaxy formation
looks much more probable.
Black holes found in small galaxies, having masses about half of the total galaxy mass surely cannot be created 
by the conventional mechanism.

The model~\cite{AD-JS}, suggested almost a quarter of century ago, very nicely fits all the new observations, 
at least qualitatively. Of course more detailed quantitative work is necessary.
A natural, though not obligatory, outcome of this model is a prediction of cosmic antimatter, mostly in the
form of compact high velocity stars in the Galaxy. The observational bounds on the abundance of such stars
are quite vague, so we may have a lot of antimatter practically at hand. Such early formed stars would behave 
as cold dark matter and so should have much larger velocity than the normal stars and hence they would populate the 
galactic halo. The mysterious MACHOs~\cite{macho} can be just such very early and almost non-luminous stars. 

An observation of antistars in the Galaxy would present a very strong support of the suggested scenario. Since antimatter
was created in clouds with the unusually high baryon-to-photon ratio $\beta$, the BBN in these clouds would
produce anomalous abundances of light elements. So it may be helpful for discovery of cosmic antimatter to
look firstly for  the clouds with anomalous chemistry in the galaxy. Though with 50\% probability it may be  
the normal matter with anomalous ${ n_B /n_\gamma}$, there is a good chance that this region is made of 
antimatter and so it should be thoroughly investigated, especially by searching for gamma radiation produced
by the matter-antimatter annihilation.

Another interesting feature of this model of antimatter formation is that heavy antinuclej, not only anti-$He^4$
can also be found in the cosmic rays.
Their abundances could be similar  to those observed in SN explosions.

\section*{Acknowledgement} This work was  supported by the grant of the Russian Federation government
11.G34.31.0047.

{}

\end{document}